\pdfoutput=1

\documentclass[11pt]{article}

\usepackage[utf8]{inputenc}
\usepackage[top=2cm, bottom=2cm, left=35mm, right=35mm]{geometry} 
\usepackage{framed}
\usepackage{amsmath}
\usepackage{amssymb}
\usepackage{graphicx}

\usepackage{txfonts}
\usepackage{titlesec}
\usepackage[superscript]{cite}
\usepackage{framed}

\titleformat{\section}{\large\bfseries\sffamily}{}{0em}{}

\begin{document}

\title{\Large\bfseries\sffamily Testing the limits of quantum mechanical superpositions$^1$}
\footnotetext[1]{This text appeared in {\sffamily Nature Physics {\bfseries 10} 271-277 (2014)}; published version differs by minor editorial changes.}
\footnotetext[2]{University of Vienna, Faculty of Physics, QuNaBioS, VCQ, Boltzmanngasse 5, 1090 Vienna, Austria.}
\footnotetext[3]{University of Duisburg-Essen, Faculty of Physics, Lotharstra{\ss}e 1, 47048 Duisburg, Germany.}

\author{Markus Arndt$^{2}$  and Klaus Hornberger$^3$}
\date{April 2014}
\maketitle
\noindent
\textbf{Quantum physics has intrigued scientists and philosophers alike, because
it challenges our notions of reality and 
locality---concepts that we have
grown to rely on in our macroscopic world. 
It is an intriguing  open question
whether the linearity of quantum mechanics extends into
the macroscopic domain.
Scientific progress over the last decades inspires hope that this
debate may be decided by table-top experiments.}

\section{Introduction}
\noindent 
The last three decades have witnessed
what has been termed \cite{Dowling2003} the second quantum revolution:
A renaissance of research on the quantum foundations,
hand in hand with growing experimental capabilities\cite{Zeilinger1999},
revived the idea of  {exploiting quantum superpositions} 
for 
technological applications, from information science \cite{Trabesinger2012,Bennett2000,Southwell2008}
to precision metrology \cite{Giovannetti2011,Riedel2010,Gross2010}.
Quantum mechanics has passed all precision tests with flying colors, but it still seems to be in conflict with our common sense. Since quantum theory knows no boundaries everything should fall under the sway of the superposition principle, including macroscopic objects. This is at the bottom of Schr{\"o}dinger's thought experiment transforming a cat into a state which strikes us as classically impossible. And yet, `Schr{\"o}dinger kittens' of entangled photons\cite{Haroche2013} and ions\cite{Wineland2013} have been realized in the lab.

So why are the objects around us never found in superpositions of 
states that would be excluded in a classical description?
One may emphasize the smallness of Planck's constant, 
or point to decoherence
theory, which describes how a system will effectively lose its quantum features when coupled to a quantum environment of sufficient 
%kh complexity\cite{Joos2003,Zurek2003}.
size\cite{Joos2003,Zurek2003}.
The formalism of decoherence, however, is based on the framework of unitary quantum mechanics,
implying that some interpretational exercise is required not to become entangled in a multitude of parallel worlds \cite{laloe2012}.
More radically, one may ask whether quantum mechanics 
breaks down beyond a certain mass or complexity scale. As will be discussed below, such ideas can be motivated by the apparent incompatibility of 
quantum theory and general relativity.
It is safe to state, in any case,  that quantum superpositions of truly massive, complex  objects  are \emph{terra incognita}.
This makes them an attractive challenge for a growing number of sophisticated experiments.

We start by reviewing several prototypical tests of the superposition principle, focusing on the quantum states of motion displayed by  material objects. Particle position and momentum variables have a 
well-defined classical analogue, and they are therefore particularly
suited to probe the macroscopic domain.
We note that aspects of macroscopicity can also be addressed in experiments with photons\cite{Fickler2012,Ma2012,Kirchmair2013}, with the phonons of ion chains\cite{Monz2011}, 
%kh
and by squeezing pseudospins 
\cite{julsgaard2001experimental,Gross2010}.

%kh
\section{State of the art}

\emph{Superconducting quantum interference devices }(SQUIDs) have recently attracted a growing interest, since they are promising
elements of quantum information processing \cite{Devoret2013}. A SQUID is a superconducting loop segmented by Josephson junctions.
Its electronic and transport properties are determined by 
a macroscopic wave function
ordering the Cooper pairs.
To exploit this 
macroscopicity it is appealing to consider a flux qubit\cite{Clarke2008} (see Figure 1A):
The single-valuedness of the wave function entails that the magnetic flux encircled by a closed-loop supercurrent must be quantized.
In particular, one can  define a symmetric and an antisymmetric linear combination of two supercurrents, which 
circulate simultaneously in opposing directions.
Billions of electrons may contribute coherently to the wave function over mesoscopic dimensions.
The difference between the clockwise and anti-clockwise currents\cite{Friedman2000_long} can reach about $2\mu$A,
amounting to a  local magnetic moment of about $10^{10}$
%kh \,$\mu_{\rm B}$. 
Bohr magnetons.
This is an impressive number, which has led to the suggestion that SQUIDs may display the most macroscopic quantum superposition to date.
However, `only' a few thousand of the Cooper pairs carrying the different currents are distinguishable\cite{Korsbakken2010}, which points to the need for an objective measure of macroscopicity (see Box).

Historically, \emph{perfect-crystal neutron quantum optics} \cite{Rauch1974} has paved the path for many interference experiments with atoms and photons.
Since the de Broglie wavelength of thermal neutrons is comparable
to the lattice constant of silicon, quantum diffraction off the nuclei may split the neutron wave function at large angles.
As of today, neutron interferometry still realizes the widest delocalization
of any massive object \cite{Zawisky2002}. With an arm separation up to 7\,cm, enclosing an area of 80\,cm$^2$,
it allows one to stick a hand between 
the two branches of a quantum state that describes a single microscopic particle (see Figure 1B).
Even though neutrons are very light neutral particles,
they are prime candidates for emergent tests of post-Newtonian gravity at short distances \cite{Nesvizhevsky2002,Jenke2011}.
With an electrical polarizability twenty orders of magnitude smaller than for atoms, neutrons are much less sensitive to electrostatic perturbations, such as charges, patch effects, or van der Waals forces.

\begin{figure}
\includegraphics[width=1.04\columnwidth]{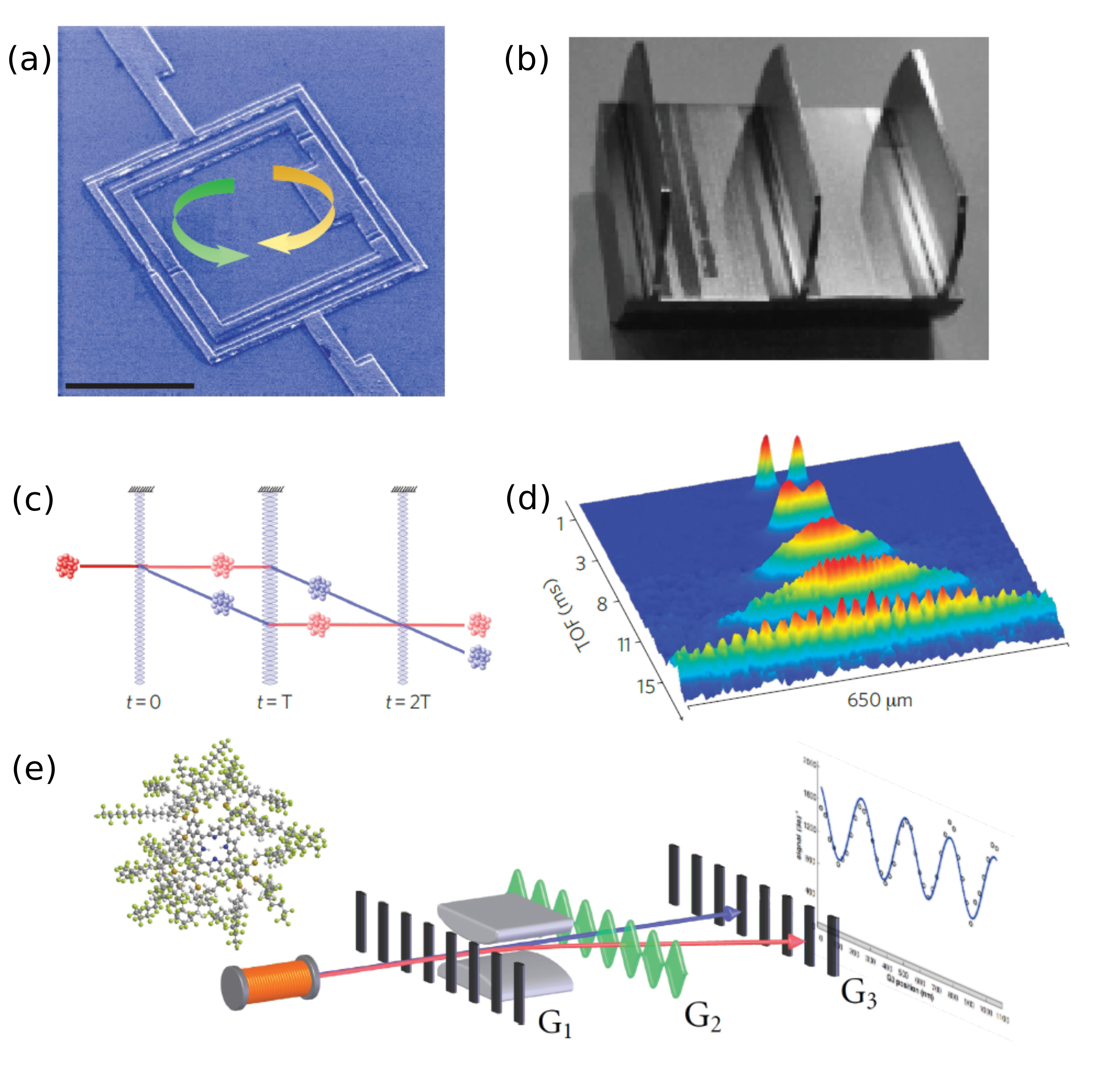}
\small{\bfseries\sffamily Figure 1---Superposition experiments.}
a) A flux qubit realizes a quantum superposition of left- and right-circulating supercurrents\cite{Friedman2000_long} with billions of 
electrons contributing to the quantum state. 
b) Neutron interferometry with perfect crystal beam splitters holds the current record in matter-wave delocalization\cite{Zawisky2002}, separating the 
quantum wave packet by up to 7\,cm.  
c) Modern atom interferometry achieves coherence times beyond two seconds with wave packet separations up to 1.5\,cm  \cite{Muentinga2013,Dickerson2013,Dimopoulos2007}.
d) Interference of two clouds of Bose-Einstein condensed diatomic Lithium molecules \cite{Kohstall2011}
%Bose-Einstein condensate of ultra-cold Rubidium atoms unite a large atomic ensemble in a single quantum state.
e) Kapitza-Dirac-Talbot-Lau interferometer for macromolecules\cite{Gerlich2007,Tuexen2010,Eibenberger2013}.
Figures reproduced with permission from: ({a}) Ref.\ \citen{Clarke2008} \copyright\  2010 NPG; ({b}) Ref.\ \citen{Zawisky2002} \copyright\ 2002 Elsevier; ({d}) displays data from Ref.\ \citen{Kohstall2011}, \copyright\ C.~Kohstall\&R.~Grimm, University of Innsbruck, Austria; ({e}) Ref.\ \citen{Tuexen2010} \copyright\ 2010 RSC.
\end{figure}

Much better control and signal to noise can be achieved by using atoms. 
\emph{Atom interferometry } (Figure 1C) started about 30 years ago \cite{Gould1986,Keith1988,Borde1989}.
The development of Raman \cite{Kasevich1991a} beam-splitters then  transformed tools of basic science into high-precision quantum sensors
which split, invert and recombine the atomic wave function in three short laser pulses (see Figure 1C).
In particular, inertial forces such as gravity and Coriolis forces\cite{Peters1999,Stockton2011} have been measured with stunning precision in experiments
that promise also new tests of general relativity \cite{Hohensee2011}.

The mass in these experiments is always limited to that of a single atom, in practice to the Cesium mass of $133$\,amu. A degree of macroscopicity can still be reached in the spatial extension of the wave function and in coherence time.
The achievable delocalization depends on the momentum transfer in the beam splitting element, while the coherence time is essentially determined by the duration of free fall
in the apparatus. Both impressively wide-angle beam splitters\cite{Mueller2008,Chiow2011} and very long coherence times \cite{Muentinga2013} have been demonstrated separately, and 
%kh ``grating'' hier unklar
%a transfer of more than 10 grating momenta 
%a multiphoton Bragg diffraction involving more than 10 photons
been recently combined in an experiment with Rubidium atoms, whose wave packets get separated for 2.3\,seconds with a maximal distance of 1.4\,centimeters \cite{Dickerson2013}.
%This allows the wave packets to  separate on the order of centimeters.
Future quantum 
%kh probes 
sensors are expected 
%kh to expand quantum coherence and 
to increase the sensitivity of quantum metrology by several orders of magnitude.
The coherence time grows only with the square root of the machine length, so that it will be practically limited to several seconds in Earth bound devices, even in high drop towers. Progress in matter-wave beam splitting will depend on improved wave-front control of the beam splitting lasers and other technology breakthroughs.
If it were possible to build interferometers of 100\,m length with beam splitters capable of transferring 
a hundred grating momenta \cite{Dimopoulos2007},
atomic matter would be delocalized over 
%kh meter-sized lengths.
distances of meters.
Even though designed for testing effects of general relativity \cite{Hohensee2011,Bouyer2010}, such experiments will also test the linearity of quantum mechanics\cite{Nimmrichter2013} as well as the homogeneity of space-time\cite{Percival1997}.

It is frequently suggested that ultra-cold atomic ensembles may serve to test the linearity of quantum physics 
%kh
even better since all atoms can be described by a joint many-body wave function once they are cooled below the phase transition to Bose-Einstein condensation (BEC, Figure 1D). 
\nocite{Kohstall2011}
Billions of non-interacting atoms may be united in a quantum degenerate state, which is however a product of single-particle states $\psi\propto(|0\rangle+|1\rangle)^{\otimes N}$, so that interference of Bose-condensed atoms depends only on the de Broglie wavelength of single atoms. A genuinely entangled many particle state $\psi\propto|0\rangle^{\otimes N}+|1\rangle^{\otimes N} $ akin to a Schr{\"o}dinger cat state would be required to reduce the fringe spacing. Such macroscopic cat states 
%kh
with regard to the particle motion
have remained an open challenge,
%kh in position space, 
even though entanglement in other degrees of freedom has been demonstrated between dozens of atoms  \cite{Riedel2010,Gross2010,Sherson2006}.
In variance to that, macromolecules and clusters open a new 
field
%kh complexity domain 
involving strongly bound
particles 
%kh in a temperature range 
with internal temperatures up to 1000\,K. When $N$ atoms
are covalently linked into a single molecule they act as a single object in quantum interference experiments. 
The entire $N$-atom system is then delocalized over two or more interferometer arms. 

Macromolecule interferometry started originally from far-field diffraction of fullerenes \cite{Arndt1999}
and works with high mass objects in currently two different settings: the Kapitza-Dirac-Tabot-Lau interferometer (KDTLI)
and an all optical interferometer in the time domain with pulsed ionization gratings (OTIMA).
Both concepts were developed and implemented at the University of Vienna \cite{Gerlich2007,Haslinger2013} and are based on similar ideas.
In high-mass 
%kh quantum 
matter wave interference we face de Broglie wavelengths between 10 femtometers and 10 picometers for objects between $10^{10}$ and $10^3$\,amu. This is more than six orders of magnitude smaller than in all experiments with ultra-cold atoms. Macromolecules are not susceptible to established laser cooling techniques, though first steps into the cavity cooling of 10$^{10}$ amu objects have been taken \cite{kiesel2013cavity, asenbaum2013cavity}.
The particles therefore start out in rather mixed states, requiring near-field interference schemes \cite{Clauser1997}. 
%kh w\"urde ich streichen, (mir&stefan) unklar was gemeint:
%Also the beam splitters must be  adjusted to  accommodate the rich set of internal states in  compound systems.
\begin{figure}
\begin{framed}
{\bfseries\sffamily Box 1---Measuring macroscopicity}
\small

How can one compare different experimental approaches for establishing large mechanical superposition states?
Various measures are on offer for attributing a size to a given state
\cite{Leggett2002,Duer2002, Bjoerk2004, Korsbakken2007, Marquardt2008, Lee2011,Froewis2012}. They presuppose a distinguished partitioning of the many particle Hilbert space into single degrees of freedom, and most of them rely on distinguished measurement or decoherence bases. Such approaches work well if the examined systems and states are of the same kind, but they do not allow us to compare disparate mechanical superposition states in an unbiased way, say superconducting ring currents with an interfering buckyball.

\hspace{1.5em}To circumvent this problem, a recent macroscopicity measure\cite{Nimmrichter2013} quantifies the empirical relevance of the concrete experiment at hand, rather than an abstract state in Hilbert space. Ultimately, any such experiment tests the hypothesis that the superposition principle is no longer valid at a certain scale.
A superposition state can thus be called the more macroscopic the better its demonstration allows one to rule out even minimal modifications of quantum mechanics that lead to classical behavior on the macroscale.

\hspace{1.5em}To turn this into a definite measure one needs to parametrize the class of minimal classicalizing modifications. This can be done without looking at specific realizations, such as the CSL model, by focusing on their observational consequences on the level of the density operator. Demanding the modification to obey basic symmetry and consistency requirements (Galilean and scale invariance, consistent treatment of identical and of uncorrelated particles) the scope of falsified theories can be characterized in the end by a single bound, a coherence time parameter $\tau_e$.
Given two experiments, the one implying a larger value of  $\tau_e$ is thus more macroscopic, and one may define its degree of macroscopicity as $\mu=\log_{10}$($\tau_e$/1\,s).
The electron is taken as reference, such that the experiment 
confirms quantum mechanics as strongly as an electron behaving
like a wave for longer than $10^{\mu}$ seconds\cite{Nimmrichter2013}.

\hspace{1.5em}The figure shows the macroscopicities for a selection of past and proposed experiments. The superconducting loop currents of Ref. \citen{Friedman2000_long} feature relatively low
due to the small electron mass and coherence time. It would be much higher in a hypothetical large SQUID with a length of 20\,mm and 1\,ms coherence time. For the oscillating micromembrane we assume that the device from Ref. \citen{Teufel2011_long} can be kept in a superposition of the zero- and  and one-phonon state for 1000 oscillation periods.

\vspace*{2ex}
\centerline{
\includegraphics[width=0.7\textwidth]{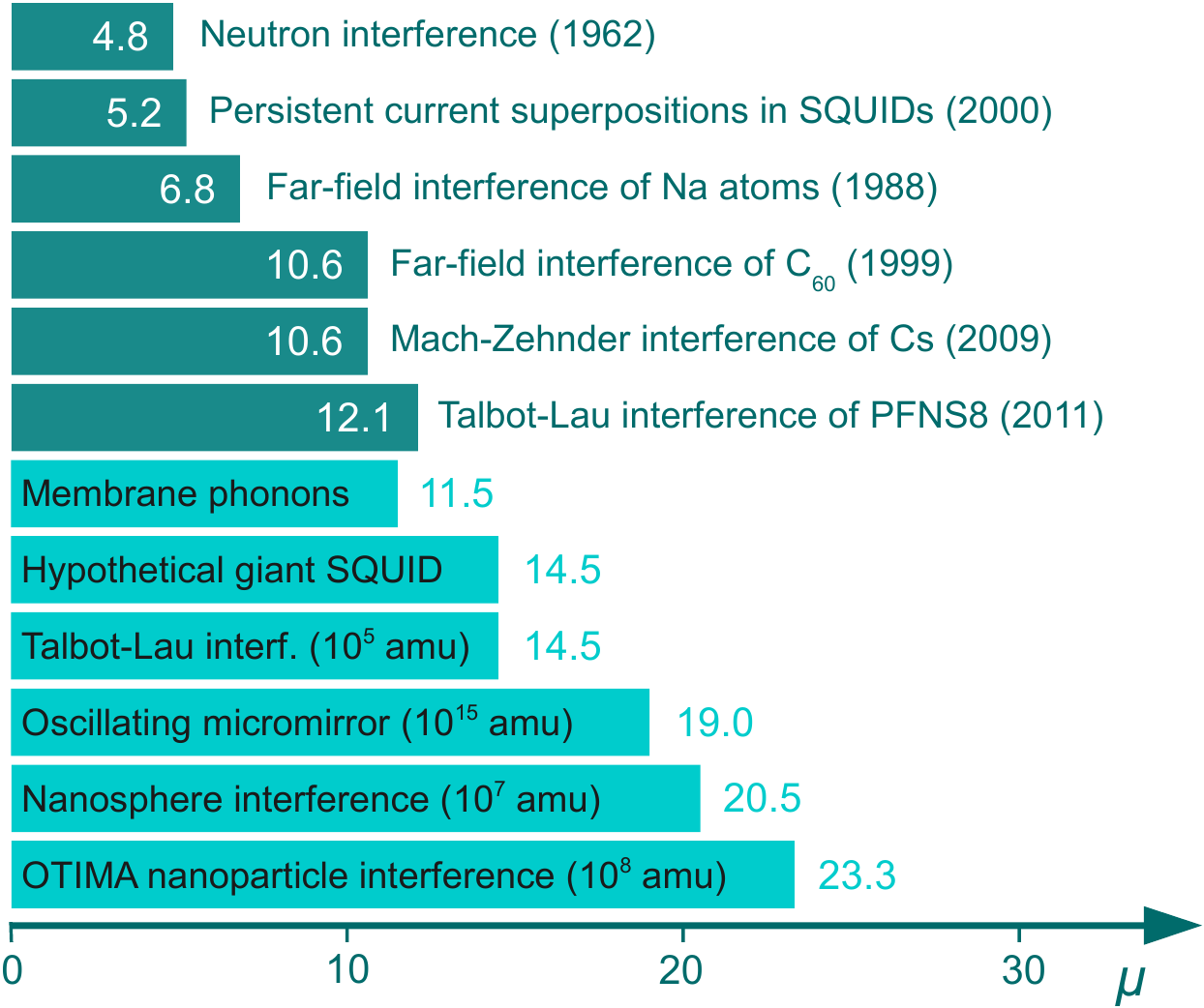}}
{\small\noindent
{\bfseries\sffamily Figure B1---Macroscopicities of different superposition experiments}
Macroscopicities $\mu$ reached in past experiments (top) and proposed tests (bottom) of the superpostion principle  as evaluated in Ref. \citen{Nimmrichter2013}.
}
\end{framed}
\end{figure}

The KDTLI interferometer is sketched in Figure 1E. It accepts a large variety of nanoparticles,
since it uses only non-resonant gratings to split (G$_1$), 
%kh rephase/
diffract (G$_2$) and probe (G$_3$) matter-waves.
%kh (Kausalitaet unklar) Since most particle sources are incoherent (thermal) 
The first grating (G$_1$) implements a spatially periodic transmission function. 
The size of the slits and the separation between G$_1$ and G$_2$ are chosen such that the position-momentum uncertainty in each
slit is sufficient to expand each particle's wave function to cover more than two slits
in G$_2$ downstream. For that G$_1$ must be an absorptive mask, here realized as a silicon nitride nanostructure.
%kh in SiN$_x$. 
Grating G$_2$, a non-resonant standing light wave, imprints a spatially periodic phase onto the matter-wave.
%kh which 
A near-field resonance effect rephases the wave functions to a molecular density pattern at the position of G$_3$.
While one might capture the emerging quantum fringe pattern on a substrate for subsequent high-resolution microscopy \cite{Juffmann2009, Juffmann2012a},
it is often convenient to scan the absorptive mask G$_3$ across the nanopattern: 
A plot of the number of transmitted particles as a function of the mask's position, reveals the molecular interferogramm (Fig 1E).

In contrast to the KDTLI, an OTIMA interferometer relies on three  pulsed  gratings which 
%kh
ionize and thus remove the molecules  at the anti-nodes of an ultraviolet standing-wave laser beam\cite{reiger2006exploration}. Such all optical gratings can handle highly polarizable or polar particles, and their pulsed nature 
allows us to profit from working in the time domain. All particles exposed to the spatially extended nanosecond 
laser pulses then see the same grating for the same time, regardless of their velocity. This eliminates numerous dispersive dephasing phenomena,
which is particularly beneficial for quantum tests at high masses \cite{Nimmrichter2011a_PRA_long,Nimmrichter2011b}.
KDTLI and OTIMA are `universal' 
%kh since 
in the sense that they can accept a wide class of different objects and
both avoid 
%kh
the detrimental effect of
van der Waals forces in G$_2$ by using non-resonant optical beam splitters.

Experiments in the KDTLI currently hold the mass record in matter-wave interference
with a functionalized tetraphenylporphyrin molecule that combines 810 atoms into one
particle with a molecular weight exceeding 10,000 amu \cite{Eibenberger2013}.
Even at an internal temperature of 500\,K this object can be 
%kh quantum
delocalized over 
%100\,
a hundred times its own diameter and for more than 1\,ms.
Very recently, the OTIMA concept has been demonstrated\cite{Haslinger2013} with clusters of molecules.
It will soon be used to explore quantum coherence at unprecedented masses \cite{Nimmrichter2011a_PRA_long}.
Both interferometers also share a high potential for quantum-assisted metrology targeting 
%kh measurements of 
internal properties, which reveal themselves even in de Broglie experiments 
%kh in interaction with 
due to the phase shift induced by
external fields\cite{Berninger2007,Gerlich2008b,Tuexen2010}.

\section{Physics beyond the Schr\"odinger equation?}

The experimental tests discussed so far confirm quantum mechanics impressively, as do high precision spectroscopic measurements \cite{PhysRevLett.84.5496,PhysRevLett.97.030801} and tests of nonlocality \cite{freedman1972,aspect1982,giustina2013}. 
Many physicists take for granted that quantum theory is valid on macroscopic scales, the more so since environmental decoherence explains why macroscopic objects \emph{seem} to assume the classically distinguished states we observe in our everyday life\cite{Joos2003,Zurek2003} (see Fig.~2).

\begin{figure}
\includegraphics[width=0.8\columnwidth]{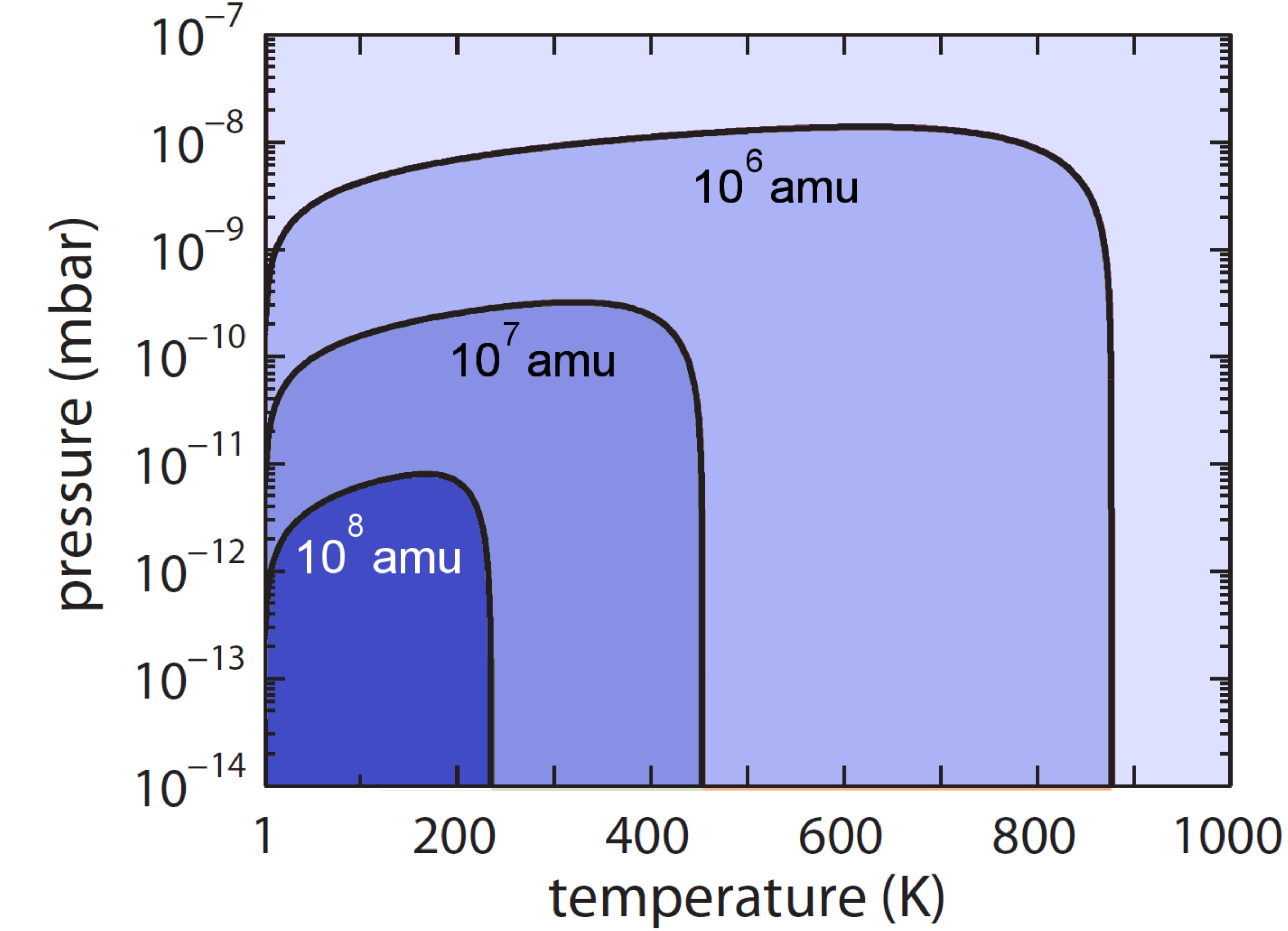}
\\
\small
{\bfseries\sffamily Figure 2---Accounting for environmental decoherence}
The theory of decoherence accounts for the impact of a quantum system on practically unobservable environmental degrees of freedom\cite{Joos2003,Zurek2003}. It can thus explain the effective super-selection of distinguished system states and the emergence of classical dynamics. From a practical point of view, decoherence theory  tells us how strongly a quantum system must be isolated from its surroundings to be still  expected to show quantum interference. The figure gives the ambient temperature and pressure requirements for observing OTIMA interference with gold clusters of 10$^6$\,amu, 10$^7$\,amu, and 10$^8$\,amu. Similarly demanding conditions for shielding environmental decoherence apply to the other described superposition tests. Figure adapted with permission from Ref. \citen{Nimmrichter2011a_PRA_long} \copyright\ 2011 APS.
\end{figure} 

Yet, there are good reasons to take seriously the possibility that
quantum theory may fail beyond some scale. A compelling one is the difficulty of reconciling quantum theory with the nonlinear laws of general relativity, which treats space-time as a dynamical entity. Most theories of quantum gravity
suggest that there is a minimal observable length scale, often associated with the Planck length. One way to account for this phenomenologically is to postulate modified commutator relations for the canonical observables,
which might be testable by monitoring the motion of massive pendulums at the quantum level \cite{Abbott2009,PhysRevLett.101.221301, PhysRevD.86.085017, pikovski2012probing, marin2012gravitational}.
The granularity of space-time might  manifest itself also
in a fundamentally non-unitary time evolution of the quantum system, which would be  observable as an intrinsic decoherence process \cite{Percival1997,PhysRevLett.93.240401,Milburn-intrinsic2006,Wang2006a}.

The alternative that gravity is not to be quantized, but fundamentally described by a classical field, suggests to extend the Schr\"odinger equation nonlinearly to account for the gravitational self-interaction\cite{Bassi2012,PhysRevLett.110.170401}.
This idea is formalized in the Schr\"odinger-Newton equation, which can be obtained as the non-relativistic limit of self-gravitating Klein-Gordon fields \cite{Giulini2012}. 
It has been hypothesized that this equation defines the time scale and the basis states of a fundamental collapse mechanism. 
%kh :
Indeed, an additional collapse-like stochastic process is required for any such non-linear extension of the Schr\"odinger equation to ensure that the time evolution maps any initial state linearly to an ensemble described by a proper density operator. Otherwise an entangled particle pair  would admit superluminal signaling, i.e. violate causality, because the nonlinearity would imprint the basis of a distant measurement onto the reduced local state \cite{gisin1989stochastic}. A gravitationally inspired non-linear modification of quantum mechanics\cite{diosi1987universal} can be made consistent with causality and observations at the price of a fictitiously large blurring of the involved mass density\cite{Bassi2012}.

The best studied nonlinear modification of quantum mechanics is the continuous spontaneous localization (CSL) model \cite{GPR1990,Bassi2003}. It augments the Schr\"odinger equation for elementary particles with a Gaussian noise term which gives rise to a continuous stochastic collapse of wave functions delocalized beyond about 100\,nm.
The origin of the stochastic process remains unspecified; one may view it either as a fundamental trait of nature, or as the repercussion of an inaccessible underlying dynamics
\cite{Adler2004}.
The CSL effect would be very weak and practically unobservable on the atomic level, but
it would get strongly amplified 
for bound atoms forming a solid, such as the pointer of a measurement device. Any superposition of macroscopically distinct positions would rapidly collapse, in agreement with Born's rule, to a `classical' state characterized by a localized, objective wave function.
This way the model serves its purpose of restoring objective classical reality on the scale of everyday objects,
allowing to dispense with the measurement postulate.

It is a contentious issue whether  such \emph{macrorealism} \cite{Leggett2002} is required in a plausible description of physical reality. Independent of that,  the CSL model 
serves as a cautionary tale.
It proves that there are competing descriptions of nature,
which predict strongly different effects at macroscopic scales, 
even though they are compatible with all experiments and cosmological observations carried out to date \cite{Feldmann2012,Bassi2012}. One may evoke metaphysical arguments in favor of one or another theory, but empirically their status is equal, and only future experiments will be able to tell them apart.

\section{Venturing towards macroscopic quantum superpositions }

Various different systems have been suggested to probe the quantum superposition principle at mesoscopic or even macroscopic scales.
This raises the question how to objectively assess the degree of macroscopicity reached in different experiments\cite{Nimmrichter2013} (see Box 1).

\begin{figure}
\centerline{
\includegraphics[width=0.66\textwidth]{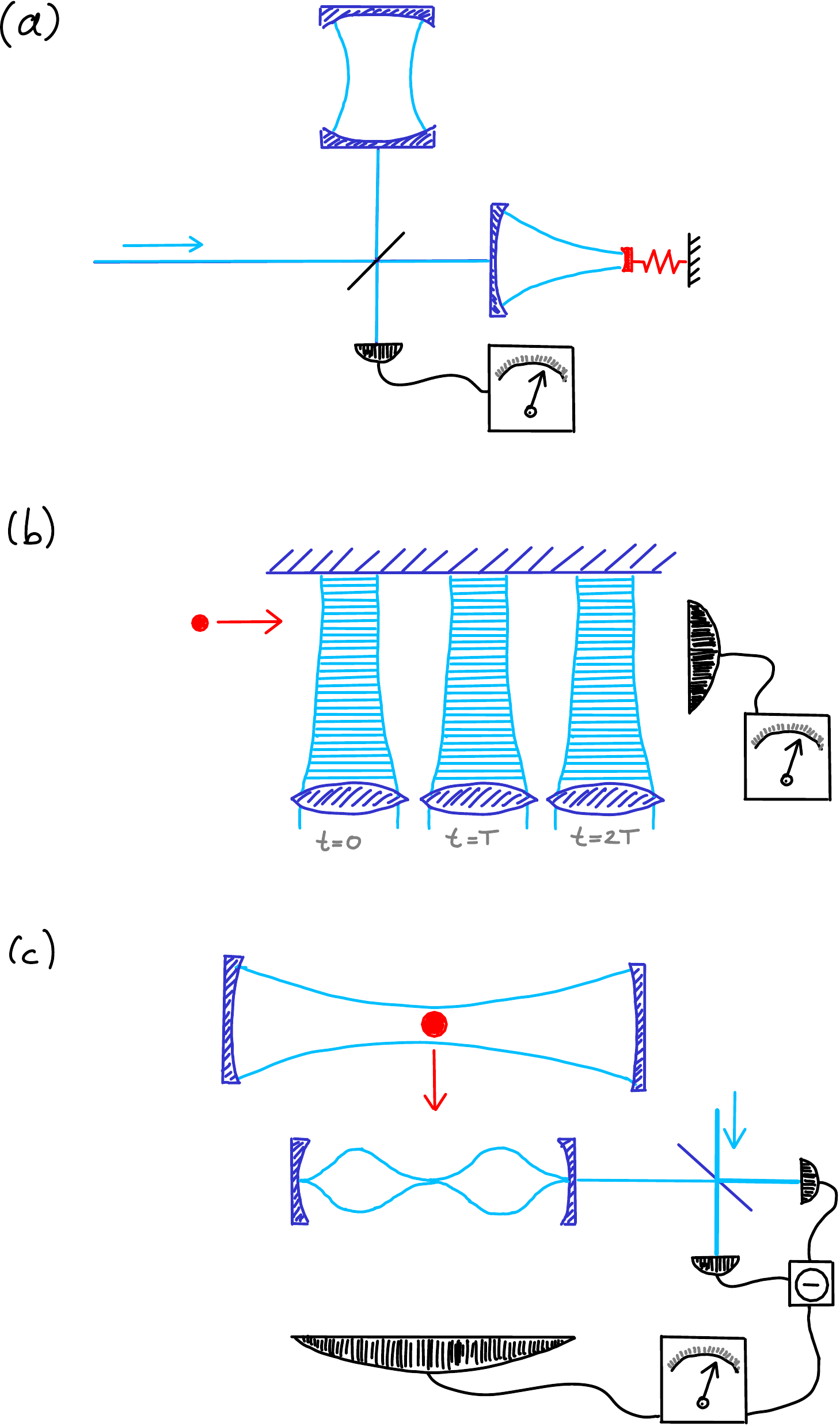}}
\small
{\bfseries\sffamily Figure 3---Interference schemes for large masses}
(a) The superposition of a micromechanical oscillator can be triggered by scattering a single photon in a Michelson interferometer.
(b) Time-domain matter wave interferometry  of nanoparticles with pulsed laser gratings is expected to be scalable to high masses.
(c) Far-field interference of nanospheres at a measurement-induced double slit may be observed by correlating the detected positions with a phase measurement. 
\end{figure} 

The  gravitational collapse hypothesis \cite{Penrose1996} inspired a proposal to create a quantum superposition in the center-of-mass motion of a micromirror \cite{Marshall2003} (Fig.~3a). A light-weight (picogram) mirror suspended from a cantilever can close a cavity acting as one arm of a Michelson interferometer.
A single photon entering the interferometer excites a superposition of the two cavity modes. 
The  radiation pressure of the single photon induces a deflective oscillation of the small mirror by about the width of the zero-point motion.
Which-path information is thus left behind once the photon escapes from the cavities, unless this occurs at a multiple of the cantilever oscillation period, when the original state of the mirror reappears.
Observing the recurrence of optical interference after one such oscillation period  would therefore prove that the mirror was in a superposition state \cite{Marshall2003,Bose1999a}.

This is a difficult experiment because a relatively massive oscillator with an eigenfrequency in the low kHz regime is required
for probing gravitational collapse.  
This implies that the oscillator ground state is reached only at micro-Kelvin temperatures. Ground state cooling is easier with lighter and more rigid MHz or GHz oscillators, and by addressing normal modes with stronger opto-mechanical coupling. This feat has been achieved recently with the flexural mode
of a circular aluminum micro-membrane using optical side-band cooling \cite{Teufel2011_long,Chan2011}.
Many groups worldwide have embarked on studying
such nanomechanical oscillators \cite{aspelmeyer2013cavity},
which can serve as an interface between quantum systems. However, it has been difficult to observe genuine quantum effects in optomechanical systems because they still lack the strong nonlinear coupling  required to generate quantum states of motion that differ qualitatively from classical ones. As a first step in this direction a piezoelectric resonator was coupled coherently to a superconducting loop \cite{OConnell2010_long}.

The distinctive feature of micromechanical devices compared to other quantum systems is their very high mass. However, the quantum delocalization of the oscillatory ground state,
%kh. 
which is a collective degree of freedom involving all the atoms,
will reach at most about one picometer in conceivable setups---a tiny fraction of the size of an atom.
This indicates why some matter-wave experiments will reach beyond the macroscopicity  of a possible superposition of the micro-membrane (see Box 1).

Since any clamped nanostructure will be prone to damping, recent proposals \cite{Chang2010,Romero-Isart2010,PhysRevA.81.023826} consider levitating dielectric nanoparticles in the focus of an intense laser beam.  
Cooling the center-of-mass motion to the ground state should be feasible, due to their lower mass and the high trap frequencies.
Moreover, the nanosphere position can be coupled nonlinearly to
a resonator light field by placing the optical trap to the node of a Fabry-P{\'e}rot cavity. This opens the possibility to create distinctively nonclassical states, and to probe the wave nature of the nano-spheres e.g. by implementing an effective double-slit\cite{Romero-Isart2011b}.
In this scheme one would simply drop the nanosphere once it has been cooled to the ground state of a dipole trap. After the wave packet is sufficiently dispersed, a laser pulse passing through a Fabry-P{\'e}rot cavity reveals the square of the position by a homodyne measurement of the cavity light field. One thus learns the distance of the sphere from the cavity center, but not whether it is on the left or right, thus effectively projecting its wave function to a spatial superposition state. An interference pattern should then be observable after a further free evolution of the sphere,  and after many repetitions, if one correlates the detected positions with the results of the homodyne measurements.
The nanosphere position would be delocalized by about the diameter of the sphere, which should be sufficiently large to test effects of the CSL collapse model.

A straightforward strategy to probe the wave nature of nanometer-sized objects is to push established matter wave interference schemes to the limits of large masses. The OTIMA interferometer should allow us to probe the quantum nature of  10$^{5}$\,amu particles if the source ejects them with  a velocity of about 10\,m/s \cite{Nimmrichter2011b}. Objects with a diameter up to 10\,nm would get delocalized over 80 nanometers. 
In the future even nanoparticles in the mass range of 10$^{8}$\,amu might be diffracted with a OTIMA scheme, e.g.\ gold clusters with a diameter of 22\,nm.  Successful interference  at these masses would falsify all current CSL predictions\cite{Nimmrichter2011a_PRA_long}.
However, it would require us to counteract the gravitational acceleration, by noise-free levitation techniques or by going to a microgravity environment,
to allow the wave function to expand over a coherence time of many seconds.
Moreover, environmental decoherence would need to be suppressed by setting the ambient pressure to below 10$^{-11}$ millibars and by cooling the apparatus to cryogenic temperatures\cite{Hornberger2012RMP}, see Fig.~2. The biggest challenge, both for OTIMA interferometry and the realization of a projective double slit, is the preparation of size-selected neutral particles in ultra-high vacuum at low internal and motional temperatures.
Some promising  first steps have been achieved by recent demonstrations of optical feedback cooling \cite{Li2011,Gieseler2012} and cavity cooling \cite{kiesel2013cavity, asenbaum2013cavity}.

\section{Perspectives}

Will the quantum superposition principle stand the test of time? We have emphasized that this question is neither crazy nor heretical. Objective modifications of quantum mechanics can be set up which agree with all observations and experiments to date, while describing a tangible breakdown of quantum theory at the macroscale. Whether quantum mechanics is universally valid is thus not an issue of conviction or metaphysical reasoning, but an empirical question, to be answered only by future experiments.

A great variety of quantum systems may be used to demonstrate mechanical superposition states, whose mass, geometric size,  and delocalization scales may vary by orders of magnitude. Any such quantum test, if carried out successfully, will rule out a generic class of objective modifications of quantum mechanics.
Using the scope of this  falsified class as a
yardstick, it is remarkable that totally different experimental approaches
lead to comparable degrees of macroscopicity (see Fig.~B1).
This suggests that there is not a single golden strategy to be pursued, and that much will depend on experimental advances and ideas.
It is thus a long and exciting journey into the realm of large quantum superpositions, and one worth taking.

\section{Acknowledgements}

We thank Stefan Nimmrichter for helpful discussions, and we
acknowledge support by the European Commission within NANOQUESTFIT (No.~304886). MA is supported by the Austrian FWF (Wittgenstein Z149-N16) and by the ERC (AdvG 320694 Probiotiqus), KH by the DFG (HO 2318/4-1 and SFB/TR12).  We thank the WE Heraeus foundation for supporting the physics school ``Exploring the Limits of the Quantum Superposition Principle''.

\end{document}